\documentclass[onecolumn,preprintnumbers,12pt,amsmath,amsssymb]{revtex4}
\usepackage{color}
\begin{document}
\title{Hawking radiation as tunneling for spherically symmetric black holes: A generalized treatment}
\author{Sudipta Sarkar}
\email{sudipta@iucaa.ernet.in}
\author{Dawood Kothawala}
\email{dawood@iucaa.ernet.in}
\affiliation{IUCAA,
Post Bag 4, Ganeshkhind, Pune - 411 007, India}

\date{\today}
\def\eq#1{{Eq.~(\ref{#1})}}



%
%
\begin{abstract}
We present a derivation of Hawking radiation through tunneling
mechanism for a general class of asymptotically flat, spherically
symmetric spacetimes. The tunneling rate $\Gamma \sim \exp{(\Delta
S)}$ arises as a consequence of the first law of thermodynamics,
$TdS=dE + PdV$. Therefore, this approach demonstrates how tunneling
is intimately connected with the first law of thermodynamics through
the principle of conservation of energy. The analysis is also
generally applicable to any reasonable theory of gravity so long as
the first law of thermodynamics for horizons holds in the form,
$TdS=dE + PdV$.
\end{abstract}
\maketitle
\noindent
\section{Introduction}
Understanding Hawking radiation is one of the key issues in any
effort to unify gravity with quantum rules. In the last few decades,
we have witnessed several independent attempts to comprehend the
fact that black holes radiates and therefore behaves like a perfect
thermal system. The original derivation of Hawking radiation
\cite{Hawking} involves the calculation of the Bogoliuyobov
coefficients between asymptotic in and out states in a collapsing
geometry. Another different approach uses the Euclidean quantum
gravity techniques \cite{euclidean}. In that case, the thermal
nature of
horizons arises from the periodicity in Euclidean time needed to get rid of the conical singularity.\\
There exists another popular and may be more physically motivated
approach for Hawking radiation based on quantum tunneling. Recently,
it has been demonstrated that such a interpretation can really work
rigorously by adopting a semiclassical method for tunneling
\cite{maulik,shrini}. The main ingredient of this work  is the
consideration of energy conservation in tunneling of a thin shell
from the hole, motivated by some earlier work \cite{wilch}. An added
advantage of tunneling approach is its simplicity and as a result,
it can be easily extended to different cases
\cite{tunneling,paddyI}. In all situations, the tunneling approach
works perfectly, giving physically meaningful results. Virtually all
the known solutions of the general relativity with horizons, have
been investigated and the results seem to be in favor of tunneling
interpretation. Despite this, \textit{there is no general approach
for the tunneling of matter from a horizon which is independent of
solution} and the main motivation of this paper is to show that
\textit{such a generalization is indeed possible}. Another important
result which comes out in this context is the relationship of
tunneling with the first law of thermodynamics for spacetime
horizons. In fact this is observed in some earlier work
\cite{1stlaw} but again, no generalized demonstration exists. Our
paper will address this issue and explore the relationship of
tunneling and first law of thermodynamics in a general spherically
symmetric setting.
\section{Formalism}
We will consider a static, spherically symmetric horizon, in an
asymptotically flat spacetime described by the metric:
\begin{equation}
ds^2 =- f(r) dt_{s}^2 + \frac{1}{f(r)} dr^2 +r^2 d\Omega^2. \label{metric}
\end{equation}
We assume this metric is the solution of field equation for a
general matter described by a energy momentum tensor
$T_{\nu}^{\mu}$. We will also assume that the function $f(r)$ has a
simple zero at $r=a$, and $f'(a)$ is finite, so that spacetime has a
horizon at $r=a$ and periodicity in Euclidean time allows us to
associate a temperature with the horizon as $T= f'(a)/4\pi$. (Even
for spacetimes with multi-horizons, this prescription is locally
valid for each horizon surface.) In order to generalize the
tunneling formalism for such a general spherically symmetric setup,
one need to tackle two key issues. The first is, how to apply energy
conservation in such a general setup, and secondly, how to write the
general form of the first law of thermodynamics in such a situation.
The first one is tricker because the concept of energy in general
relativity is difficult to define. One has to depend on some
heuristic arguments to apply energy conservation for a general
setup. The answer to the second issue already exists in a different
approach to understand the dynamics of gravity and thermodynamics
\cite{paddy}. Following that methodology, it is possible to show
that in a general spherically symmetric background the Einstein's
equations evaluated on the horizons can be cast into a form
\begin{eqnarray}
TdS= dE_h +PdV \label{EHthermo}
\end{eqnarray}
where $S$ is the entropy associated with the horizon at $r=a$, $E_h$
is its energy and $P$ is the radial pressure which is equal to
$T_{r}^{r}(a)$. The expression of energy $E_h$ is equal to $a/2$
which is the irreducible mass of the hole ($E_h$ is also equal to
the Misner-Sharp energy of the horizon) and $V = 4 \pi a^3 / 3$ is
the relevant volume for our analysis. In spite of superficial
similarity, Eq.~(\ref{EHthermo}) is different from the conventional
first law of black hole thermodynamics \cite{Hawking2}, due to the
presence of $PdV$ term. The validity of this expression can be seen,
for example, in the case of Reissner-Nordstrom black hole for which
$P \neq 0$. If a chargeless particle of mass $dM$ is dropped into a
Reissner-Nordstrom blackhole, then an elementary calculation shows
that the energy defined above as $E_h \equiv  a/2$ changes by $
dE_h= (da/2) =(1/2)[a/(a-M)]dM \neq dM$ while it is $dE_h+PdV$ which
is precisely equal to $dM$ making sure $TdS=dM$. So we need the
$PdV$ term to get $TdS=dM$ when a \textit{chargeless} particle is
dropped into a Reissner-Nordstrom blackhole. More generally, if $da$
arises due to changes $dM$ and $dQ$, it is easy to show  that
Eq.~(\ref{EHthermo}) gives $TdS=dM -(Q/a)dQ$ where the second term
arises from the electrostatic contribution from the horizon surface
charge. Dynamically, Eq.~(\ref{EHthermo}) is best interpreted as the
energy balance under infinitesimal virtual displacements of the
horizon normal to itself and therefore must be linked with energy
conservation and thus to the tunneling process.

This result can be formally interpreted by noting that in standard
thermodynamics, we consider two equilibrium states of a system
differing infinitesimally in the extensive variables like entropy,
energy, and volume by $dS$, $dE_h$ and $dV$ while having same values
for the intensive variables like temperature $(T)$ and pressure
$(P)$. Then, the first law of thermodynamics asserts that $TdS = PdV
+ dE_h$ for these states. In a similar way, Eq.~(\ref{EHthermo}) can
be interpreted as a connection between two quasi-static equilibrium
states where both of them are spherically symmetric solutions of
Einstein equations with the radius of horizon differing by $da$
while having same source $T_{\mu\nu}$ and temperature $T = f'(a) / 4
\pi $. This formalism does not care what causes the change of the
horizon radius and therefore much more formal and generally
applicable. More recently, this approach has been extended to the
general class of Lovelock  gravity\cite{sudipta}, stationary Kerr
like and time dependent spherically symmetric spacetimes
\cite{dawood} as well as for the FRW class of metrics \cite{cai}.
This approach suggests that the relevant energy of the horizon which
enters into the first law of thermodynamics under the change of the
horizon radius is $E_h = a/2$. Since, first law of thermodynamics is
basically a statement of the conservation of energy, $E_h = a/2$
should be the appropriate notion of energy of the horizon in the
tunneling process. Now, to describe across-horizon phenomena like
tunneling, it is necessary to choose coordinates which, unlike
Schwarzschild like coordinates, are not singular at the horizon. A
particularly suitable choice is obtained by introducing Painlev\'e
type of coordinates where the metric is given by,
\begin{eqnarray}
ds^2 = -f dt^2 \pm 2\sqrt{1-f}dt dr +dr^2+r^2 d\Omega^2\label{metricP}
\end{eqnarray}
where the plus (minus) sign denotes the spacetime line element of
the out-going (incoming) particles across the horizon. We want to
consider the tunneling of matter across the horizon. By treating the
matter as a sort of De-Broglie S-wave, its trajectory can be
approximately determined as \cite{trajec,shrini},
\begin{eqnarray}
\dot{r} = \pm \frac{f}{2\sqrt{1-f}}\approx \pm \kappa (r-a)\label{trajectory}
\end{eqnarray}
where $\kappa$ is the surface gravity of the horizon at $r=a$. The
imaginary part of the action for an s-wave outgoing positive energy
particle which crosses the horizon outwards from $r_{i}$ to $r_{f}$
can be expressed as
\begin{equation}
{\rm Im}~ {\cal S} = {\rm Im} \int_{r_{i}}^{r_{f}} p_r \, dr = {\rm
Im} \int_{r_{i}}^{r_{f}} \! \! \int_0^{p_r} d p'_r \, dr \; .
\label{integrals}
\end{equation}
We multiply and divide the integrand by the
two sides of Hamilton's equation
$\dot{r} = +\left. {d{\cal H} \over d p_r} \right | _r \,$,
change variable from momentum to
energy, and switch the order of integration to obtain
\begin{equation}
{\rm Im}~ {\cal S} = {\rm Im} \int \!\! \int_{r_i}^{r_f} {dr \over
\dot{r}} \, d{\cal H}
\end{equation}
Now, using \eq{trajectory} and performing the radial integral with
appropriate contour prescription, one immediately finds that, for
outgoing case \cite{maulik},
\begin{equation}
{\rm Im}~ {\cal S} = -\int \frac{d{\cal H}}{2 T}, \label{action}
\end{equation}
where $T= \kappa /2 \pi$, the Hawking temperature associated with the horizon. The next task is to find the correct
 Hamiltonian to evaluate this integral.\\
For this, we appeal to energy conservation to guess the form of ${\cal H}$.\\
We first note that since there is no explicit time dependence, the
Hamiltonian should be equal to the total energy of the system. To
find an expression for the total energy, we again note that, the
horizon separates the spacetime into two parts, which can be
labelled as inside and outside. Our previous arguments \cite{paddy}
show that the energy associated with the horizon is $E_h = a/2$.
Also, for matter with $T_{\nu}^{\mu}$ as the energy momentum tensor,
the contribution from the matter field in the outside region should
be,
\begin{eqnarray}
E_m = -\int_{\Sigma}T_{\nu}^{\mu} \xi^{\nu} d\Sigma_{\mu},
\end{eqnarray}
where the integration is over the 3-surface $\Sigma$ which extends
from the horizon to infinity. (The negative sign merely reflects the
fact that in our convention $-T_{t}^{t}$ is the energy density of
the matter.) Hence the total energy of the spacetime should be,
\begin{eqnarray}
E_T &=& E_h + E_m \\ \nonumber
&=& \frac{a}{2} -\int_{\Sigma}T_{\nu}^{\mu} \xi^{\nu} d\Sigma_{\mu}.
\end{eqnarray}
For our particular case, $\xi^{\nu} = (1,0,0,0)$. Then the energy expression reduces to,
\begin{eqnarray}
E_T &=&\frac{a}{2} -\int_{r=a}^{\infty}T_{t}^{t} 4 \pi r^2 dr  \label{energy}\\
&=&\frac{a}{2} -\int_{r=a}^{\infty}T_{r}^{r} 4 \pi r^2 dr.\nonumber
\end{eqnarray}
In the second step, we have used the fact that, for the metric in \eq{metricP}, $G_{t}^{t} =G_{r}^{r}$.

Now, in order to show that the expression of $E_T$ is indeed the
energy of the spacetime, we evaluate it for two separate cases.
First, for the trivial case of Schwarzschild black hole for which
$T_{\nu}^{\mu} = 0$ and $a = 2M$, which gives $E_T = M$ as desired.
Next, for the Reissner-Nordstrom black hole, we have,
\begin{eqnarray}
T_{\nu}^{\mu} = \frac{Q^2}{8 \pi r^4} (-1,-1,1,1)~~\textrm{and}~~E_h = M -\frac{Q^2}{2 a}.
\end{eqnarray}
It is possible to provide a simple explanation of the energy
expression $E_h$. Consider any general $r > a $ and then we can
write, for Reissner-Nordstrom black hole,
\begin{eqnarray}
 1- \frac{2M}{r} +\frac{Q^2}{r^2} =1- \frac{2 \mu(r)}{r}, ~~\textrm{where}~~\mu(r)=M -\frac{Q^2}{2 r}.
\end{eqnarray}
Hence, the quantity $\mu (r) = M -Q^2/2r$ can be interpreted as the
mass energy inside the sphere of radius $r$ and for $r=a$, we have
$\mu (a) = E_h$ validating the case for $E_h$ as the energy of the
horizon. Substituting this into \eq{energy}, it is easy to show that
even in this case $E_T = M$.

As a result of tunneling across the horizon, some matter either
tunnels out or in across the horizon. Hence, the parameters which
fix the radius of the horizon ( mass, charge etc.) change and that
leads to a change in the radius of the horizon. In fact, \textit{the
only physical change occurring due to the process of tunneling is in
the radius of the horizon}. We suppose, due to tunneling the radius
changes from $a$ to $a + \delta a$ (note that $\delta a$ is positive
for incoming shell and negative for outgoing shell). Both $E_h$ and
$E_m$ and as a result $E_T$ depends of $a$. Therefore, let the
energy $E_T$ changes to $E^{i}_{T}(a)$ to $E^{f}_{T}(a+\delta a)$
and the difference $E^{f}_{T} - E^{i}_{T}$ can be attributed to the
shell. So, by energy conservation, we can immediately write,
\begin{eqnarray}
d{\cal H} &=&E^{f}_{T}(a+\delta a) -E^{i}_{T}(a)  \label{Derivation}\\ \nonumber
&=& \frac{\delta a}{2} - \left(\int_{a+\delta a}^{\infty} -\int_{a}^{\infty}\right)T_{r}^{r} 4 \pi r^2 dr.\\ \nonumber
&=&\frac{\delta a}{2} +T_{r}^{r}(a)~4 \pi a^2 \delta a\\ \nonumber
&=& dE_h + PdV.
\end{eqnarray}
Substituting this in \eq{action}, we ultimately get,
\begin{eqnarray}
{\rm Im}~ {\cal S} = -\int \frac{dE_h + PdV}{2 T} =- \int \frac{dS}{2},
\end{eqnarray}
where we have used the first law of thermodynamics as in \eq{EHthermo}.

Written in this form, one can also see an immediate generalization
of this approach, at least in spherical symmetry, to any theory of
gravity for which a suitable generalization of $a/2$, the
Misner-Sharp energy, can be motivated (note that the $PdV$ term
generalizes in a {\it natural} way, i.e., with $dV$ interpreted as
the areal volume). Indeed, such an expression for energy exists for
the Lanczos-Lovelock lagrangians \cite{sudipta,entropygb}. Replacing
$a/2$ with that expression, one can explicitly show that we get the
correct result; specifically, we obtain the correct scaling of
entropy (which is no longer proportional to area) for the
Lanczos-Lovelock lagrangians.

Another interesting point which comes in this regard is that, entire
analysis is totally local. In fact, all contributions from the
spatial infinity cancels out in the second step of \eq{Derivation}.
Hence, in principle this approach is extendable to spacetimes having
multiple horizons. Now, the semi classical tunneling rate is given
by,
\begin{eqnarray}
\Gamma \sim  e^{- 2 {\rm Im}~ {\cal S}} = e^{ \int_{S_i}^{S_f} dS} = e^{\Delta S},
\end{eqnarray}
where $\Delta S = S_f - S_i $. This is the well-known result
obtained in \cite{maulik} for a general, asymptotically flat,
spherically symmetric
background.\\
Now, we would like to discuss the case for the class of metrics in
which $g_{\,rr}g_{\,t_s t_s} \neq -1$. The spacetime metric is given
by,
\begin{equation}
ds^2 =- f(r) dt_{s}^2 + \frac{1}{g(r)} dr^2 +r^2 d\Omega^2. \label{metricG}
\end{equation}
The horizon is given by a simple zero of the function $f(r)$ at
$r=a$. Temperature associated with this horizon can be shown to be
$T=\sqrt{f'(a)g'(a)}/4\pi$. The fact that $r=a$ is a null surface,
and requiring the regularity of Ricci scalar at $r=a$, further
impose two conditions on $g(r)$ at $r=a$. (a) $g(a)=0$ and (b)
$f'(a)=g'(a)$. Using these, the temperature associated with the
horizon at $r=a$ becomes $T=g'(a)/4\pi$. Using these conditions, and
writing the metric in the Painleve form, one can show that
\cite{dawood, visser} the energy momentum tensor, {\it on the
horizon}, again has the form,
\begin{eqnarray}
T_{t}^{t}|_{r=a} = T_{r}^{r}|_{r=a} ; ~~~T_{\theta}^{\theta}|_{r=a} = T_{\phi}^{\phi}|_{r=a}. \label{emten}
\end{eqnarray}
(where, as earlier, $t$ is the Painleve time coordinate.) It can be
shown that \cite{paddy} the relevant energy of the horizon, $E_h$,
is again $a/2$, so that the first equality in Eq.~(\ref{energy})
still holds. Proceeding as before, using energy conservation, we can
write,
\begin{eqnarray}
d{\cal H} &=&E^{f}_{T}(a+\delta a) -E^{i}_{T}(a) \\ \nonumber
&=& \frac{\delta a}{2} - \left(\int_{a+\delta a}^{\infty} -\int_{a}^{\infty}\right)T_{t}^{t} 4 \pi r^2 dr.\\ \nonumber
&=&\frac{\delta a}{2} +T_{t}^{t}(a)~4 \pi a^2 \delta a\\ \nonumber
&=&\frac{\delta a}{2} +T_{r}^{r}(a)~4 \pi a^2 \delta a\\ \nonumber
&=& dE_h + PdV.
\end{eqnarray}
Where we have used Eq.~(\ref{emten}) in the fourth line. The
remaining steps are identical as before, and give the same tunneling
rate, $\Gamma \sim \exp{\Delta S}$.

Hence, we find that, for a general spherically symmetric case, the
derivation of this result requires only local physics. There was
neither an appeal to Euclideanization nor any need to invoke an
explicit collapse phase. The simple facts that tunneling changes the
horizon radius and energy conservation, are enough to find the semi
classical tunneling rate.

Another striking feature which naturally appears from this analysis
is the relationship between tunneling and the first law of
thermodynamics. The tunneling rate is ultimately found to be,
\begin{eqnarray}
\Gamma &\sim& e^{\int \frac{dE_h + PdV}{T}}\\ \nonumber
&\sim& e^{\Delta S}
\end{eqnarray}
In order to obtain the second expression from the first, we have to
apply the generalized first law $TdS =dE_h + PdV$. This fact clearly
shows that tunneling interpretation of black hole radiation is
intimately related with the first law of thermodynamics. This is
also expected because the basic input behind both tunneling and the
first law is same, namely the conservation of energy.

We would also like to point out the importance of the assumption of
spherical symmetry in our calculation. The two main ingredients
which depend on this assumption are,
\begin{enumerate}
\item The fact that $G_{t}^{t} =G_{r}^{r}$ on the horizon, which ultimately allows us to write the $PdV$ term.
This equality can be seen as a result of the equality of $g_{\,rr}$ and $g_{\,tt}$, and their first derivatives,
 on the horizon, for the metric \eq{metricG}. But, it is also possible to show that this result is valid even for
  non spherically symmetric case, near the Killing horizon of any stationary but nonstatic (and nonextremal) black
  hole spacetime \cite{DBH}. In that case, the near horizon structure of the Einstein tensor, and hence that of
  $T^{\mu}_{\nu}$, admits a block diagonal representation, in which $T_{t}^{t} =T_{r}^{r}$ (with the time coordinate
  interpreted appropriately). Hence, this assumption is not specific to spherical symmetry.
\item The quantities $dE_h$ and $PdV$ have a natural interpretation in spherical symmetry and it is difficult to
construct analogous quantities in non spherically symmetric case. But, at least for spacetimes with a timelike killing
vector, there is a notion of volume \cite{maulikV} and it may help to generalize our arguments for non spherical situation.
\end{enumerate}
But the fact that this generalized approach works for spherical symmetry is intriguing enough, and it may be possible
 that this can also be extended to non spherically symmetric situations, by interpreting various quantities appropriately.
\section{Conclusion}
Our analysis generalizes the tunneling approach of Hawking radiation
for a general spherically symmetric and asymptotically flat set up.
The basic inputs that go into the calculation is the energy function
in \eq{energy} and application of energy conservation. With these
two inputs and also using \eq{EHthermo}, one immediately recovers
the results of \cite{maulik} in this general context. This shows the
validity of tunneling process for a general spherically symmetric
horizon and also its relation with the first law of thermodynamics.
In fact in our analysis, the result $\Gamma \sim \exp{(\Delta S)}$
comes out as a natural consequence of the first law $TdS =dE_h +
PdV$, and therefore suggests an intimate dependence on each other
through the principle of conservation of energy.

\textit{Note:} While completing this work, we came across the
reference \cite{tun}, which also reaches the same conclusion
regarding the relationship of tunneling and the first law of
thermodynamics through a totally different approach. In fact, unlike
\cite{tun}, our method is \textit{independent of the theory of
gravity and makes no assumption of entropy being proportional to
area}. Another crucial difference is the presence of the $PdV$ term
in our analysis. Hence, the approach presented here is more
generally applicable as long as the first law of thermodynamics
holds in the form $TdS=dE_h + PdV$.
\section{Acknowledgments}
The authors thank Maulik Parikh and T. Padmanabhan for constructive
comments and criticisms on the earlier drafts. We would also like to
thank Naresh Dadhich for useful discussions. Both SS and DK are
supported by the Council of Scientific \& Industrial Research,
India.

\end{document}